\documentclass[sigconf,nonacm,authorversion]{acmart}
\usepackage{microtype}
\usepackage{xurl}
\usepackage[colorinlistoftodos]{todonotes} 
\usepackage{tikz}
\usepackage{caption}
\usetikzlibrary{positioning, fit, arrows.meta, backgrounds, shapes.geometric, shapes.misc}
\usepackage{booktabs}
\usepackage{pgfplots}                                            \pgfplotsset{compat=1.18}
\usepackage{array}
\usepackage{comment}
\usepackage{xspace}
\usepackage{afterpage}

\newcommand{\bench}{PAGE\xspace}

\newcommand{\AI}{A\hspace{0.05em}I}
\AtBeginDocument{%
  }

\setcopyright{none}
\copyrightyear{2026}
\acmYear{2026}

\begin{document}

\title{Persona Guardrail: A Production-Grade Defense Framework for Agentic Systems}

\author{Bijeeta Pal}
\author{Sridhar Reddy Maddireddy}
\author{Muhaimin Bin Munir}
\authornote{University of Texas at Dallas; work done during an internship at Uber.}
\author{Zoltan Puha}
\authornote{Uber, Netherlands.}
\author{Max Zhurovich}
\author{Adi Raghavendra}
\author{Sean Tout}
\affiliation[obeypunctuation=true]{%
  \institution{Uber,}
  \country{USA\\[\smallskipamount]
    \normalfont\texttt{\{palb, smaddireddy, muhaimin, zoltan.puha, mzhurovich, adi.raghavendra, tout\}@uber.com}}}








\renewcommand{\shortauthors}{Pal et al.}

\begin{abstract}
Large language model-based agents are increasingly deployed to perform domain-specific tasks by interacting with enterprise knowledge, tools, and external services. Existing runtime guardrails primarily target prompt injection and other attack-specific behaviors under a black-box threat model, but provide limited guarantees that agents operate within their intended functionality. As a result, production agents remain vulnerable to malicious requests and out-of-domain queries that existing defenses often fail to distinguish.

We present Persona Guardrail, a production-grade runtime defense framework that enforces explicit functional boundaries for customer-facing agentic \AI{} systems through synchronous input and output validation driven by semantic allowlist and blocklist specifications. We also introduce PAGE (Persona-Aware Guardrail Evaluation), a benchmark for evaluating function-specific guardrails across benign, adversarial, and out-of-domain interactions on both user and agent turns. Compared with a generic LLM-based guardrail, Persona Guardrail improves overall accuracy from 85.7\% to 95.9\%, increases out-of-domain detection from 57.3\% to 93.5\%, and reduces the false-approved rate from 25.0\% to 4.7\%. Currently deployed in production, Persona Guardrail meets its latency budget while sustaining a very low false-block and false-allow rate under realistic production workloads. These results demonstrate that Persona Guardrail provides a practical, scalable, and production-ready foundation for securing agentic AI systems.
\end{abstract}



\maketitle

\section{Introduction} 



Recent advances in large language models (LLMs) have significantly improved their capabilities in reasoning, planning, tool use, and decision-making ~\cite{yao2023react,schick2023toolformer}, enabling them to autonomously execute complex workflows that were previously achievable only through carefully engineered software systems. These advances have enabled the emergence of what is called agentic \AI{} systems—LLM-powered applications ~\cite{wu2024autogen} that can autonomously interact with tools, APIs, knowledge bases, and software services to accomplish a broad spectrum of user goals. By combining the  capabilities of LLMs with traditional software components, the agentic systems provide a powerful building block for a diverse set of applications, ranging from ride booking and customer support automation to transaction management, information retrieval, and personalized assistance.

As agentic \AI{} systems become increasingly integrated into production environments, ensuring their security has emerged as a critical challenge ~\cite{owasp2025llm, kim2026agenticsok}. Unlike traditional software systems, agentic applications process natural-language inputs while simultaneously possessing the ability to perform privileged actions in production. This combination introduces new attack surfaces that do not exist in conventional applications. Recent research has identified a broad spectrum of attacks against agentic systems, including prompt injection, indirect prompt injection through untrusted external content, tool poisoning, retrieval-augmented generation poisoning, agent goal hijacking etc ~\cite{debenedetti2025agentdojo, greshake2023prompt, jia2025taskshield}. However, these attacks assume different threat models and require different defense mechanisms.

In this paper, we focus on the practical threat model encountered by many production-facing agents, where a malicious end user has black-box access to the agent through its public interface and 
cannot manipulate the agent's internal tools, retrieved documents, memory, or execution environment. In this setting, the adversary attempts to induce the agent to perform actions outside its intended functionality, bypass safety policies, or produce harmful responses through carefully crafted user queries. Consequently, indirect prompt injection, tool poisoning, and memory poisoning are outside the evaluated threat model. Our solution may still block a policy-violating final response caused by such attacks, but the evaluated guardrails do not inspect intermediate actions.

Existing research on agentic \AI{} security has primarily focused on two complementary directions. The first seeks to develop defenses against specific classes of attacks, such as prompt injection attacks ~\cite{promptshield, jia2025taskshield}. The second evaluates the effectiveness of adversarial attack strategies against LLM-based agents and measures their impact on system behavior ~\cite{agentharm2024, debenedetti2025agentdojo}. More recently, systematization-of-knowledge (SoK) \cite{kim2026agenticsok} studies have consolidated these attacks and defenses into comprehensive frameworks, providing a structured understanding of the emerging security landscape of agentic \AI{} systems. While these efforts have significantly advanced our understanding of agentic \AI{} security, they largely focus on attack taxonomies, threat models, and isolated defense mechanisms. Comparatively little attention has been paid to the practical challenges of deploying comprehensive security solutions in large-scale production environments, where latency, scalability, reliability, and operational complexity are often as important as security.

In this paper, we address this gap by presenting \textit{Persona Guardrail}, a production-grade defense framework designed for large-scale agentic systems. Persona Guardrail is based on a simple yet powerful principle: every agent should operate within a well-defined functional boundary that specifies its intended capabilities, permissible actions, and expected behaviors. The framework continuously enforces these boundaries by validating both inputs to the agent and outputs generated by the agent, ensuring that interactions remain consistent with the agent's intended functionality even in the presence of adversarial inputs.

Our work differs from prior research in its emphasis on practical deployment. Persona Guardrail is designed for the realistic production threat model in which external users interact with customer-facing agents through a black-box interface. The framework operates under real-world constraints, including stringent latency requirements, high request volumes, and evolving adversarial behavior. Rather than focusing on a single attack class, Persona Guardrail provides a unified runtime defense that enforces an agent's functional boundaries, mitigating malicious requests while 
filtering benign but out-of-domain queries. This enables production agents to remain responsive, secure, and aligned with their intended functionality under real-world deployment conditions.

To the best of our knowledge, this is the first work to present and deploy a comprehensive, production-grade runtime defense framework for a large-scale, customer-facing agentic system, and to evaluate its effectiveness and operational feasibility under real-world deployment constraints.
This paper makes the following contributions:
\begin{itemize}
\item \textbf{Production-Scale Security Framework}. We present Persona Guardrail that enforces explicit functional boundaries for customer-facing agentic \AI{} systems through input and output validation.
\item \textbf{Functional Boundary Enforcement}. We introduce a principled approach for defining and enforcing agent-specific functional boundaries, enabling the detection and prevention of behaviors that deviate from an agent's intended purpose.
\item \textbf{Guardrail Evaluation Benchmark}. We introduce PAGE, the first benchmark that systematically evaluates runtime guard\-rails 
for function-specific agents across both user requests and agent responses, spanning benign, adversarial, and out-of-domain interactions.
\item \textbf{Comprehensive Evaluation}. We empirically demonstrate the effectiveness, scalability, and operational feasibility of Persona Guardrail on representative production agents under realistic workloads.
\item \textbf{Practical and Scalable Deployment}. We describe the architectural and operational considerations required to deploy security guardrails at scale while satisfying strict latency and reliability requirements.

\end{itemize}

\section{System Overview and Problem Setting}
In this work, we investigate the design of a comprehensive defense framework for agentic \AI{} systems. We begin by describing the system architecture, problem setting, design assumptions, and threat model that motivate the proposed defense mechanism.

\subsection{System Setting}

We consider a generic architecture representative of modern production scale agentic \AI{} systems. The system receives user requests in the form of natural-language instructions, structured inputs (e.g., JSON), or a combination of both. In addition to user-provided information, the frontend application may attach contextual metadata (e.g., user state, session information, or application-specific context) that assists the agent in decision making. Before reaching the agent, every incoming request is processed by the input Persona Guardrail, which validates the request against the agent's functional boundaries and security policies. Persona Guardrail can block or send through the request to the agentic system, which comprises one or more large language models (LLMs), enterprise knowledge bases, retrieval components, internal and external tools, APIs, and other software services. During task execution, the agent may perform multiple reasoning steps and interact with these components to retrieve information, invoke tools, or execute actions before producing a final response. The generated response is subsequently processed by the output Persona Guardrail, which validates it against the agent's intended functionality and security policies before it is returned to the user.

\subsection{Threat Model}

We consider an adversarial setting in which an external malicious user has unrestricted query access to the agentic system and can observe all responses returned by the system. The adversary possesses general knowledge of the agent's intended functionality, public interfaces, and expected behavior, but does not have access to the system's internal state, including system prompts, intermediate reasoning, memory, tool interactions, orchestration logic, or communication between internal components. We assume the underlying LLMs, infrastructure, and trusted software components are uncompromised. Within this threat model, the adversary aims to manipulate the agent through carefully crafted inputs to achieve one or more of the following objectives:

1. \textbf{Instruction Manipulation and Prompt Extraction.} The adversary attempts to extract confidential system prompts, reveal internal instructions or policies, or manipulate the agent into violating its intended behavior through attacks such as prompt injections.

2. \textbf{Generation of Harmful or Policy-Violating Responses.} The adversary seeks to induce the agent to produce responses or perform actions that violate organizational policies, safety requirements, or legal and regulatory constraints, potentially resulting in reputational, financial, or legal harm to the organization.

3. \textbf{Functional Boundary Violation.} The adversary attempts to cause the agent to operate outside its intended scope by eliciting responses or executing actions unrelated to its designated functionality. For example, an attacker may attempt to coerce a ride-booking assistant into providing medical advice or revealing internal operational information, thereby violating the agent's predefined functional boundaries.

Our objective is to design a defense framework that mitigates these attacks while preserving the utility, responsiveness, and intended functionality of the agent under normal user interactions.

\subsection{Performance Requirements}

Persona Guardrail is designed to operate under the stringent latency requirements of production agentic systems. Both the input and output guardrails make blocking decisions—requests must be validated before agent execution, and responses must be validated before being returned to the end user. Hence, the guardrails cannot be executed asynchronously with the agent. As a result, the primary design objective is to achieve robust security while incurring only a small and predictable latency overhead that preserves the responsiveness required by production-scale deployments.
\subsection{Potential Approaches}

From a high-level perspective, an agentic guardrail can be viewed as a reasoning system that determines whether an input or output lies within the functional and security boundaries of an agent. A straightforward approach is to leverage a general-purpose LLM by providing it with a natural-language specification of the agent's intended functionality, permissible actions, and security policies, and asking it to classify each interaction accordingly. While conceptually simple, this approach presents several practical challenges for production deployment.

First, a comprehensive natural-language specification is difficult to construct and maintain as the underlying agent evolves. Production agents frequently undergo updates, including changes to prompts, tools, APIs, retrieval sources, and supported capabilities. Consequently, the guardrail specification must evolve alongside the agent to accurately reflect its intended behavior. Second, invoking a medium- or large-scale reasoning model for every request introduces significant latency. Since both the input and output guardrails operate synchronously on the critical execution path, their latency directly impacts end-to-end response time and user experience. Finally, a static guardrail is unlikely to remain effective in dynamic production environments where both adversarial attack strategies and legitimate user behavior continuously evolve.

These challenges raise several fundamental design questions. How can the functional and security specification of an agent be maintained as the system evolves? How should guardrail performance be measured, particularly in terms of false positive and false negative rates? How can production misclassifications be systematically incorporated to refine the guardrail over time? More broadly, how can a guardrail continuously adapt to emerging attack techniques and distribution shifts in user requests while maintaining low latency and high reliability?

Existing defenses only partially address these challenges. Rule-based and policy-based guardrails~\cite{rebedea2023nemo, inan2023llamaguard} provide deterministic enforcement but are difficult to scale across heterogeneous agents and often fail to capture context-dependent or previously unseen attacks. Model-based guardrails~\cite{promptshield, jia2025taskshield} offer greater flexibility and generalization but typically rely on expensive reasoning models and have received relatively little attention regarding continuous adaptation, deployment at production-scale, or systematic evaluation under evolving workloads. More generally, recent systematization efforts~\cite{kim2026agenticsok} categorize existing defenses into runtime protection, secure-by-design mechanisms, and component hardening, while identifying the need for practical, adaptive defenses that can operate under real-world deployment constraints.

Motivated by these challenges, we design Persona Guardrail as an adaptive, production-oriented runtime defense framework. Rather than relying solely on static rules or expensive reasoning, Persona Guardrail combines lightweight inference with a closed-loop feedback mechanism that continuously refines its functional and security boundaries using signals collected from production traffic. This design enables the framework to adapt to both emerging adversarial attack patterns and shifts in user behavior while satisfying the stringent latency and scalability requirements of production agentic systems.

\section{Persona Guardrail Design}
Persona Guardrail is a production-grade defense framework designed to ensure that an agentic system operates within its intended functional and security boundaries. The framework consists of three primary components: 1) \textit{Functional Boundary Specification}, 2) \textit{Input Guardrail Enforcement}, and 3) \textit{Output Guardrail Enforcement}.


The key insight behind Persona Guardrail is that every agent has a well-defined purpose. Rather than attempting to detect every possible attack in isolation, Persona Guardrail first models the intended functional boundary of the agent and then enforces this specification throughout the agent's interaction lifecycle. Both the input and output guardrails are responsible for ensuring that the interaction remains consistent with the agent's functional boundary.

\subsection{Functional Boundary Specification}

The central abstraction underlying Persona Guardrail is the \textit{functional boundary} of an agent. Intuitively, the functional boundary characterizes the set of user interactions that are consistent with the agent's intended functionality, security requirements, and product specifications.

\paragraph{Definition 1 (Functional Boundary)} Let $\mathcal{Q}$ denote the space of all possible user queries submitted to an agent. The functional boundary partitions $\mathcal{Q}$ into two disjoint subsets:

[
$\mathcal{Q} =
\mathcal{Q}_{\text{allow}}
\cup
\mathcal{Q}_{\text{block}},$
] where

\begin{itemize}
    \item $\mathcal{Q}_{\text{allow}}$ denotes the set of queries that are consistent with the intended functionality of the agent and should be processed.
    \item $\mathcal{Q}_{\text{block}}$ denotes the set of queries that should be rejected: malicious or unsafe queries, together with out-of-domain queries in deployments configured to reject them.
\end{itemize}

The objective of Persona Guardrail is to approximate this partition to minimize both false positives (rejecting legitimate queries) and false negatives (allowing malicious or out-of-domain queries). Since the behavior of production agents and their users continuously evolves, the functional boundary is not static; instead, it is continuously refined using production telemetry, attack simulations, and feedback collected from deployment.

In practice, explicitly enumerating the sets $\mathcal{Q}_{\text{allow}}$ and $\mathcal{Q}_{\text{block}}$ is infeasible because natural language admits an effectively unbounded space of expressions. Persona Guardrail therefore represents the functional boundary using semantic descriptions that capture both the intended functionality of the agent and security constraints.

\subsubsection{Hybrid Functional Boundary Representation}

We investigate three approaches for representing the functional boundary:

\begin{itemize}
    \item Allowlist-only
    \item Blocklist-only
    \item Hybrid allowlist-blocklist
\end{itemize}

An \textit{allowlist-only} representation specifies the categories of requests that the agent is permitted to process. While this approach provides strong security, it often becomes overly conservative in practice. Because it is impossible to exhaustively enumerate every legitimate user request, incomplete allowlists frequently result in elevated false positive rates and poor user experience.

Conversely, a \textit{blocklist-only} representation specifies the set of queries
that the agent should reject. These include both security-driven and
functionality-driven constraints. Security-driven entries correspond to known
adversarial behaviors, such as prompt injection, jailbreaks, and other attacks
that attempt to manipulate the agent or violate organizational policies.
Functionality-driven entries correspond to out-of-domain (OOD) requests that fall
outside the intended purpose of the agent. Although such requests may be entirely
benign, processing them can result in poor user experience, hallucinations, or
unintended behavior. Whether out-of-domain queries are included in the blocklist
is configurable and depends on the deployment requirements of the agent. Some
applications are designed to operate as specialized assistants and therefore
reject requests outside their intended functionality, while others may choose to
answer or gracefully redirect out-of-domain requests. Persona Guardrail supports
both deployment models by allowing agent owners to configure the desired
functional boundary.


To reduce ambiguity at inference time, the allowlist and blocklist should
  overlap as little as possible. Where the two specifications intersect, they
  present the classifier with conflicting evidence and raise classification
  uncertainty. Persona Guardrail therefore identifies overlapping entries
  between the two lists and adjudicates them during guardrail construction and
  refinement, so that each capability falls under a single, unambiguous rule.

\subsubsection{Constructing the Functional Boundary}

The functional boundary is constructed from multiple complementary sources.

\textit{Agent implementation and product specifications.} During initial deployment, the allowlist is generated by analyzing both the implementation of the agent and its product requirements. We employ a code-oriented LLM to analyze the agent implementation and infer the classes of user requests that the agent is designed to support. This analysis considers the agent's available tools, APIs, retrieval components, system prompts, and external resources to produce an initial semantic specification of its intended functionality.

\textit{Production traffic analysis.} Following deployment, Persona Guard\-rail 
operates in a shadow mode during which it observes production traffic without enforcing decisions. Legitimate interactions that are not adequately represented in the initial specification are incorporated into the allowlist, enabling the functional boundary to evolve alongside changing user behavior and product capabilities.

\textit{Security-driven blocklist generation.} The initial blocklist is
constructed from two complementary sources. First, it incorporates known
adversarial techniques targeting agentic \AI{} systems, including prompt injection,
indirect prompt injection, jailbreaks, goal hijacking, functional-boundary
violations, and tool misuse. Second, it captures classes of out-of-domain
queries that lie outside the intended scope of the agent. As new attacks emerge
and user interaction patterns evolve, the blocklist is continuously refined
through a feedback loop, allowing the framework to adapt to both evolving
adversarial threats and changing product requirements.

By combining implementation analysis, production observations, and continuously
evolving security knowledge, Persona Guardrail maintains an adaptive
approximation of the agent's functional boundary that remains effective under
both evolving adversarial threats and shifts in legitimate user behavior.
Persona Guardrail enforces the functional boundary at two stages of the agent
execution pipeline. Before the request reaches the agent, the input guardrail
determines whether the query lies within the functional boundary. After the
agent generates a response, the output guardrail verifies that the generated
response remains consistent with the same functional boundary. The following
sections describe these enforcement mechanisms in detail.

\subsection{Input Guardrail Enforcement}

The input guardrail enforces the functional boundary before a request reaches
the agentic system. Its objective is to determine whether an incoming user
request should be processed by the agent or rejected. The input allowlist
specifies the classes of user requests that fall within the intended
capabilities of the agent, whereas the input blocklist captures malicious
inputs, known adversarial attack patterns, and, optionally, out-of-domain
requests that the agent is not designed to handle. To enforce the input
functional boundary, we investigated four alternative architectures:

\noindent1) \textit{Embedding-based Retrieval.} The incoming query is embedded and compared against representative allowlist and blocklist examples using vector similarity. The query is classified based on its nearest neighbors and a calibrated similarity threshold.

\noindent2) \textit{LLM-based Classification.} A low-latency reasoning model receives the user query together with the functional boundary specification and directly classifies the query as \textit{allow} or  \textit{block}.

\noindent3) \textit{Hybrid Retrieval and LLM.} Queries exhibiting very high semantic similarity to existing allowlist or blocklist examples are classified using the embedding model. Queries that cannot be classified with sufficient confidence are forwarded to the LLM classifier for semantic reasoning.

\noindent4) \textit{Fine-tuned Classification Model.} A model is fine-tuned using agent-specific allowlist and blocklist examples to directly predict the appropriate classification.

We evaluated these architectures along four dimensions: \textit{classification performance, inference latency, operational complexity, and generalization across heterogeneous agents.}
The embedding-based approach provides the lowest inference latency but requires careful calibration of similarity thresholds and continuous maintenance of an embedding database containing representative allowlist and blocklist examples for every agent. As agents evolve, maintaining these examples becomes increasingly expensive, and the approach exhibits limited robustness to semantically novel requests. The fine-tuned classifier performs well when evaluated on agents represented in its training data. However, its performance degrades when deployed on previously unseen agents or when the functional boundary evolves, necessitating periodic retraining and increasing operational complexity. The hybrid architecture improves classification accuracy by combining fast retrieval with LLM reasoning, but introduces additional infrastructure complexity and requires maintaining both an embedding index and an LLM-based inference pipeline.

Based on these trade-offs, we adopt an \textbf{LLM-based classifier} using \textbf{Qwen3-30B-A3B}~\cite{qwen3vl2025} as the input guardrail. We find that this approach offers the best overall trade-off across these dimensions: low operational complexity, low inference latency, strong classification performance, and generalization across diverse agentic systems. Moreover, because the functional boundary is represented semantically rather than as agent-specific examples, the same model can be deployed across multiple agents with minimal customization while adapting naturally to evolving boundary specifications.

\subsection{Output Guardrail Enforcement}

The output guardrail enforces the functional boundary before the agent's response is returned to the user. Its objective is to verify that the generated response remains consistent with the agent's intended functionality, organizational policies, and security requirements. The output allowlist specifies the expected characteristics and permissible actions of the agent's responses, while the output blocklist captures prohibited behaviors, including policy violations, unsafe or harmful responses, disclosure of confidential information, hallucinated capabilities and other outputs that violate the agent's functional or security boundaries.

As with the input guardrail, we evaluated the same four architectures (embedding-based retrieval, LLM-based classification, hybrid retrieval–LLM, and a fine-tuned classifier). We adopt the same LLM-based classifier (Qwen3-30B-A3B)~\cite{qwen3vl2025} for output enforcement, as it provides the best balance between inference latency, deployment simplicity, and generalization across diverse agents. Furthermore, using a common classification model for both input and output enforcement simplifies deployment and maintenance while enabling a consistent semantic interpretation of the functional boundary throughout the agent execution pipeline. The output guardrail serves as the final enforcement point before the response reaches the end user and may block a policy-violating response even when an attack has influenced the agent. 

\begin{table}[t]
\centering
\caption{Comparison of candidate architectures for input and output guardrail enforcement.}
\label{tab:guardrail-architectures}
\resizebox{\linewidth}{!}{%
\begin{tabular}{lcccc}
\toprule
\textbf{Approach} &
\textbf{Latency} &
\textbf{Classification} &
\textbf{Cross-Agent} &
\textbf{Operational} \\
&
&
\textbf{Performance} &
\textbf{Generalization} &
\textbf{Complexity} \\
\midrule

Embedding &
$\checkmark\checkmark\checkmark$ &
Low &
Low &
High \\

Fine-tuned &
$\checkmark\checkmark$ &
Medium &
Medium &
Medium \\

Hybrid &
$\checkmark\checkmark$ &
High &
High &
High \\

LLM-based (Qwen3-30B-A3B) &
$\checkmark\checkmark$ &
High &
High &
Low \\

\bottomrule
\end{tabular}%
}
\end{table}

\subsection{Guardrail Model Selection}
\label{sec:model-selection}

Model selection for Persona Guardrail is driven by both guardrail performance and production latency. To ensure a responsive user experience, we require the model's end-to-end P90 inference latency. This requirement restricts the set of models that can be practically deployed.

We initially experimented with both closed- and open-weight models. Closed models, including models served through OpenAI and Anthropic APIs, require requests to be sent to externally hosted inference
endpoints. In our experiments, the resulting network and serving overhead prevented these models from consistently satisfying our sub-100\,ms latency requirement. We therefore focused on open-weight
models that could be deployed within our own serving infrastructure, allowing us to control inference placement, caching, hardware configuration, and other serving optimizations.

Among locally deployable models, we evaluated Qwen2.5-3B, Qwen2.5-7B, Qwen3-8B, and Qwen3-30B-A3B. We additionally evaluated Qwen3-VL-30B as a higher-capacity alternative. Each model was evaluated on the same sets of benign, out-of-domain (OOD), and adversarial queries for a representative production agent. The benign evaluation contains three subsets representing different production traffic distributions: most-frequent queries, randomly sampled queries, and an extensive benign set. For OOD and adversarial datssets, we report the percentage of
queries incorrectly approved; therefore, lower values indicate stronger rejection. Table~\ref{tab:model_selection} summarizes the results.

\begin{table*}[t]
\centering
\caption{Comparison of locally deployable models for Persona Guardrail.
Values report the percentage of queries approved by each model. Higher
is better for benign (Approve) datasets, while lower is better for OOD
and adversarial datasets. Latency reports median inference latency.}
\label{tab:model_selection}
\small
\resizebox{\textwidth}{!}{%
\begin{tabular}{lccccc}
\toprule
\textbf{Dataset / Metric} &
\textbf{Qwen3-VL-30B} &
\textbf{Qwen3-30B-A3B (8-bit)} &
\textbf{Qwen3-8B (8-bit)} &
\textbf{Qwen2.5-7B} &
\textbf{Qwen2.5-3B} \\
&
&
\textbf{(Deployed)} &
&
& \\
\midrule
Approve -- Most Frequent $\uparrow$ & 99\% & \textbf{100\%} & 92\% & 84\% & 49\% \\
Approve -- Random $\uparrow$        & 96\% & \textbf{99\%}  & 93\% & 83\% & 67\% \\
Approve -- Extensive $\uparrow$     & 96\% & \textbf{99\%}  & 85\% & 77\% & 53\% \\
\midrule
OOD -- Generic $\downarrow$          & 1\%  & 5\%  & \textbf{0\%} & 0.5\% & 5\% \\
Adversarial -- Generic $\downarrow$  & \textbf{7\%} & 15\% & \textbf{7\%} & 12\% & 11\% \\
\midrule
Median latency $\downarrow$          & 150\,ms & 95\,ms & 170\,ms & 168\,ms & \textbf{81\,ms} \\
\bottomrule
\end{tabular}%
}
\end{table*}

The results expose an important trade-off between benign utility, security, and latency. The smaller models can achieve competitive rejection rates on OOD and adversarial requests, but at the cost of substantially higher false blocking of legitimate traffic. For example, Qwen2.5-3B has the lowest median latency (81\,ms) and approves only
18\% of targeted adversarial requests, but approves just 49--67\% of the benign evaluation sets. Qwen2.5-7B and Qwen3-8B improve benign utility, but still approve only 77--93\% of benign requests and exhibit
median latencies of 168\,ms and 170\,ms, respectively.

Qwen3-30B-A3B provides a substantially better utility-latency trade-off. It approves 99--100\% of benign requests across all three benign evaluation sets while maintaining a median inference latency of
95\,ms. Although some smaller models achieve lower false-approval rates on individual OOD or adversarial sets, their substantially higher benign false-block rates make them unsuitable for our production setting, where both security and preservation of legitimate agent functionality are important.

Qwen3-VL-30B provides the closest benign classification performance, approving 96--99\% of legitimate requests, and achieves lower false-approval rates than Qwen3-30B-A3B on the evaluated OOD and adversarial sets. However, its median inference latency is 150\,ms, exceeding our latency target. In contrast, Qwen3-30B-A3B uses a mixture-of-experts architecture with approximately 30B total parameters while activating only approximately 3B parameters per token. This architecture provides sufficient model capacity for semantic functional-boundary classification while keeping inference cost
compatible with synchronous guardrail enforcement.

Based on these results, we select Qwen3-30B-A3B as the shared classifier for both the input and output Persona Guardrails. Among the evaluated models, it provides the best overall trade-off between preserving benign utility, rejecting functional-boundary violations, and satisfying the production latency constraint. Using the same model for both guardrails also simplifies serving and maintenance and allows agent-specific behavior to be expressed through functional-boundary specifications rather than separate model deployments.


\section{Experimental Evaluation}
We evaluate Persona Guardrail on \textbf{PAGE (Persona-Aware Guardrail
Evaluation)} (Section~\ref{sec:page}), our persona-scoped benchmark spanning
four AgentDojo~\cite{debenedetti2025agentdojo} domains---Workspace, Slack,
Banking, and Travel (Table~\ref{tab:agentdojo_domains})---under a malicious-user
threat model in which an adversary interacts directly with the agent endpoint.
Because PAGE labels both \emph{user} turns (\textsc{pre-app}, evaluated by the
input guardrail) and \emph{agent} turns (\textsc{post-app}, evaluated by the
output guardrail), a single benchmark exercises both guardrails of the
framework. Every configuration is evaluated on identical PAGE splits so that
performance differences are attributable to the guardrail configuration alone.

Table~\ref{tab:agentdojo_domains} summarizes the four domains, their in-scope
function, and representative off-persona actions a malicious user might
at\-tempt---the behaviours the guardrail must reject. We take only the domain
definitions from AgentDojo; we do not run its environment, tools, or
indirect-injection attack suite. The results therefore do not measure
end-to-end prevention of indirect injection or unauthorized intermediate
actions.

\begin{table}[t]
\caption{Agent domains in PAGE, their supported capabilities, and representative prohibited actions.}
\label{tab:agentdojo_domains}
\centering
\small
\resizebox{\columnwidth}{!}{%
\begin{tabular}{@{}lll@{}}
\toprule
\textbf{Domain} &
\textbf{Capabilities} &
\textbf{Representative prohibited actions} \\
\midrule
Workspace &
Email, calendar, cloud drive &
Export mail/files; access another user's mailbox \\
Slack &
Messages, channels, shared links &
Forward private direct messages; leak workspace data \\
Banking &
Balances, payments, transfers &
Transfer funds to unauthorized accounts; reveal credentials \\
Travel &
Flights, hotels, restaurants, cars &
Share itineraries or personal IDs; modify another user's booking \\
\bottomrule
\end{tabular}%
}
\end{table}

\subsection{The Benchmark}
\label{sec:page}

Existing public agent benchmarks are designed to measure task utility, harmful behavior, or robustness against specific attacks, such as indirect prompt injection, and therefore do not
provide the labels a persona guardrail needs: whether a \emph{user turn}
(before the agent acts) or an \emph{agent turn} (before a response is returned)
falls inside or outside the agent's declared functional boundary. To evaluate
Persona Guardrail under its intended malicious-user threat model, we construct \textbf{PAGE (Persona-Aware Guardrail Evaluation)}, a purpose-built benchmark in which
every message is (i) attributed to one of the
four AgentDojo personas, (ii) tagged with the turn on which the guardrail fires
(\textsc{pre-app}/user or \textsc{post-app}/agent), and (iii) labelled with a
\emph{request kind}---\textit{benign}, \textit{adversarial},
\textit{out-of-domain} (OOD), or \textit{mixed}. Every label is grounded in the
same natural-language allowlist that instantiates the guardrail, so the dataset
and the enforced policy share a single source of truth. In total, PAGE
comprises $2{,}180$ labelled messages ($2{,}076$ synthesized around the four
personas plus $104$ projected from AgentHarm ~\cite{agentharm2024}) and is, to our knowledge, the
first guardrail benchmark that jointly covers user and agent turns across four
production domains with benign, out-of-domain, and adversarial intents. We
release it to support reproducible evaluation of functional-boundary
enforcement.

\subsubsection{Generation Methodology}
\label{sec:data_gen}

The dataset was produced by a human-guided, semi-automated pipeline that
combines four complementary sources, chosen so that no single generator biases
the corpus and so that the evaluated classifier never sees text
authored by itself.  Table~\ref{tab:data_provenance} summarizes the provenance of the
$2{,}076$ persona-centric messages.

\noindent1) \textbf{Human-authored seeds
(AgentDojo~\cite{debenedetti2025agentdojo}).}
The verbatim user tasks of each AgentDojo suite provide realistic \emph{benign}
in-domain requests, while the suite's injection-task goals provide grounded
\emph{adversarial} intents. These anchor the corpus in human-written,
domain-authentic language.

\noindent 2) \textbf{LLM synthesis.} A frontier instruction-tuned model
(Claude Opus~4.8 with reasoning), \emph{distinct} from the guardrail classifier
under test, expands the benign set and generates off-persona ``hallucination''
content (e.g., poems, code, opinions) for agent turns. Using a different model
family for generation and evaluation avoids self-evaluation bias.
    
\noindent3) \textbf{Deterministic adversarial templates.} Attack turns are
    synthesized from parameterised templates rather than an LLM. The frontier
    model refused to author attack payloads, and templates additionally give
    exact reproducibility and, through slot filling, broad lexical variety while
    keeping each attack grounded in an AgentDojo~\cite{debenedetti2025agentdojo}/AgentHarm~\cite{agentharm2024} pattern. Slots draw
    from $9$ exfiltration addresses, $6$ external sites, $5$ adversarial IBANs,
    $8$ third-party identities, $7$ authority personas, $7$ cities, $5$ venues,
    and $6$ monetary amounts, combined with $8$ user and $5$ agent phrasing
    prefixes.
    
\noindent 4) \textbf{Cross-persona projection.} A request that is perfectly valid
    for persona~$A$ is, by construction, a functional-boundary violation for
    persona~$B$. Projecting benign messages across personas yields $288$
    \emph{hard} OOD cases that are fluent and in-distribution yet out of scope,
    which are far more challenging than generic off-topic text.

    \emph{Mixed} cases are then assembled programmatically by grafting an
out-of-scope or adversarial rider onto a benign in-scope carrier (e.g., a valid
booking request followed by an injected exfiltration instruction), exercising
the guardrail's behaviour when legitimate and illegitimate intents co-occur in a
single turn.

\begin{table}[t]
\centering
\small
\caption{Provenance of the persona-centric synthetic corpus ($n=2{,}076$).}
\label{tab:data_provenance}
\resizebox{\columnwidth}{!}{%
\begin{tabular}{lcl}
\toprule
\textbf{Source} & \textbf{\#} & \textbf{Role in the corpus} \\
\midrule
AgentDojo (verbatim/adapted)      & 234 & benign + adversarial seeds \\
LLM synthesis (Claude~Opus~4.8)     & 972 & benign + OOD hallucinations \\
Deterministic templates           & 582 & adversarial + mixed \\
Cross-persona projection          & 288 & hard out-of-domain \\
\midrule
\textbf{Total}                    & \textbf{2{,}076} & \\
\bottomrule
\end{tabular}%
}
\end{table}

\subsubsection{Composition and Coverage}
\label{sec:data_composition}

The corpus is balanced across the four personas ($519$ messages each) and spans
both guardrail turns ($1{,}280$ user, $796$ agent). Request kinds are
distributed as $1{,}004$ benign, $448$ adversarial, $444$ out-of-domain, and
$180$ mixed, giving $1{,}184$ \emph{allowed} and $892$ \emph{should-block}
references. 
The AgentHarm projection adds $104$ malicious
user requests ($52$ adversarial, $52$ OOD) spanning eight harm categories
(Fraud, Cybercrime, Harassment, Copyright, Sexual content, Disinformation, Hate,
and Drugs), mapped onto the four personas to test whether functional-boundary
enforcement generalises to externally sourced misuse.

\subsubsection{Diversity Analysis}
\label{sec:data_diversity}

A central goal of the dataset is breadth: we want each request kind to span the
range of phenomena a production guardrail encounters, so that strong performance
is evidence of genuine functional-boundary understanding rather than overfitting
to a narrow template.

\noindent\textbf{Adversarial coverage.} Attack turns instantiate six distinct
attack families (Table~\ref{tab:attack_taxonomy}), covering data exfiltration,
access to third-party data, sending data or funds to unauthorized destinations,
account privilege and configuration tampering, direct prompt injection, and
social-engineering via authority impersonation. Each family is realised for all
four personas on both user and agent turns, and the slot-filled templates
(Section~\ref{sec:data_gen}) yield many surface realisations per family, so the
guardrail cannot succeed by memorising a fixed set of trigger phrases.

\begin{table}[t]
\centering
\small
\caption{Attack families among adversarial turns, with representative intents.
Counts are for template-generated user turns; agent turns mirror the same
families as compromised confirmations, and $46$ additional cases are AgentDojo
injection goals.}
\label{tab:attack_taxonomy}
\resizebox{\columnwidth}{!}{
\begin{tabular}{llc}
\toprule
\textbf{Attack family} & \textbf{Representative intent} & \textbf{\#} \\
\midrule
Unauthorized recipient   & send data/funds to an unprovided address & 56 \\
Third-party data access   & read/act on another user's data          & 49 \\
Social engineering        & authority impersonation to extract data   & 46 \\
Data exfiltration         & bulk export to an external sink           & 44 \\
Prompt injection          & ``ignore instructions'' / system override & 33 \\
Privilege/config tampering& change email, payees, passwords, admins   & 29 \\
\bottomrule
\end{tabular}
}
\end{table}

\noindent\textbf{Out-of-domain coverage.} OOD messages combine three families:
(i)~\emph{cross-persona} violations ($288$ cases; a request valid for another
agent, projected roughly evenly across all four source personas), which are the
hardest because they are fluent and task-like; (ii)~\emph{generic} off-scope
user queries (general knowledge, trivia, lifestyle); and (iii)~off-persona
\emph{agent hallucinations} (poems, code, opinions) that test the
\textsc{post-app} boundary. This mixture ensures the guardrail is measured on
both obviously-off-topic text and the subtler case of competent-but-out-of-scope
requests.

\noindent\textbf{Benign coverage.} The $1{,}004$ benign messages combine $168$
human-authored AgentDojo tasks with $836$ LLM-synthesized variants spanning both
user requests and agent responses, so the ``allow'' class reflects a wide range
of legitimate phrasings, multi-step tasks, and confirmation styles rather than a
handful of canonical prompts. Overall the corpus contains $1{,}725$ unique
messages ($83.1\%$ of the total) drawn from a vocabulary of $3{,}222$ word types.

\noindent\textbf{Length and structural variation.} Message length varies widely
and systematically with request kind (Table in Appendix~\ref{sec:appendix}). Terse
single-clause attack commands (median $14$ words) sit alongside long multi-step
benign itineraries and cross-persona OOD requests (up to $121$ words), while
mixed turns are longest (median $37$, up to $152$ words) because they concatenate
a benign carrier with an injected rider. This $5$--$152$ word span
forces the guardrail to reason over both short
imperatives and long, distraction-laden inputs.

Taken together---four balanced personas, dual guardrail turns, four request
kinds, six attack families, three OOD families, and a wide, kind-dependent
length distribution over a $3{,}222$-word vocabulary ---these properties make
PAGE a broad and reusable benchmark for persona-scoped guardrails, rather
than a narrow test tailored to a single defense.

\noindent\textbf{Metrics.} We treat \emph{block} as the positive class: the guardrail
should block off-persona content and approve in-scope content. For each
configuration we report the four decision outcomes---\emph{correct approved}
(TN) and \emph{correct blocked} (TP) versus \emph{false approved} (FN, a
should-block turn wrongly allowed) and \emph{false blocked} (FP, a legitimate
turn wrongly rejected)---together with the derived \emph{false-approved rate}
($\mathrm{FAR}=\mathrm{FN}/(\mathrm{FN}+\mathrm{TP})$) and \emph{false-blocked
rate} ($\mathrm{FBR}=\mathrm{FP}/(\mathrm{FP}+\mathrm{TN})$), and standard
accuracy, precision, recall, and F1. 

\subsection{Guardrail Configurations}
\label{sec:configs}

We study a progression of configurations of increasing specificity, all
evaluated on the same \bench splits:

    \noindent 1) \textbf{Baseline (LLM judge).} The classifier receives only a generic
    natural-language description of the agent and is asked to label each turn as
    benign, adversarial, or out-of-domain. No allowlist, blocklist, or persona
    is supplied. This isolates the value added by an explicit functional
    boundary. 

   \noindent 2) \textbf{+ Allowed intents (allowlist).} We supply the agent's persona
    \emph{allowlist}---the enumerated set of permitted user and agent
    intents---and admit a turn only when it matches an allowed intent.

   \noindent 3) \textbf{+ Iterated allowlist.} We refine the allowlist by studying the
    \emph{false-approved} and \emph{false-blocked} turns of the previous
    configuration and sharpening the intent wording accordingly (tightening
    over-broad intents and clarifying scope boundaries). This is a single
    data-driven refinement pass.

   \noindent 4) \textbf{Blocklist only.} As a counterpoint, the persona is expressed
    as a \emph{blocklist} of disallowed intents, with no allowlist.

\noindent 5) \textbf{Allowlist + blocklist.} The persona combines both an allowlist
    and a blocklist.

The last two configurations are exploratory: whether an explicit blocklist
improves over the iterated allowlist is an empirical question, and we let the
measured results (Section~\ref{sec:overall_results}) determine the narrative.

\subsection{Overall Detection Performance}
\label{sec:overall_results}

Table~\ref{tab:overall_rates} reports aggregate performance across
configurations and Table~\ref{tab:overall_counts} the underlying decision
counts. The comparison isolates the value of the functional boundary. The
baseline judge---given only a generic agent description---attains reasonable
adversarial detection ($92.6\%$) but cannot enforce a functional boundary: it
catches only $57.3\%$ of out-of-domain turns, giving a $25.0\%$ false-approved
rate and $85.7\%$ accuracy. Introducing the persona allowlist raises
out-of-domain detection to $83.3\%$ and cuts the false-approved rate to
$9.9\%$; the single iteration pass---driven entirely by inspecting the
false-approved and false-blocked turns---raises out-of-domain detection further
to $93.5\%$ and reduces the false-approved rate to $4.7\%$, lifting accuracy to
$95.9\%$; a second refinement pass and a held-out, multi-split validation of
this iterative gain are reported in Appendix~\ref{sec:appendix}
(Table~\ref{tab:iteration_heldout}). Crucially, the benign false-blocked rate stays essentially constant
across all three configurations ($\approx 3.7\%$), so these security gains do
not cost utility.

The blocklist-only and allowlist+blocklist configurations settle whether an
explicit blocklist helps. A blocklist \emph{alone} is far weaker: it reaches
only $71.2\%$ accuracy because, although it catches $87.2\%$ of adversarial
turns, it flags just $14.5\%$ of out-of-domain requests---a blocklist
enumerates attacks, not the agent's function, so it cannot recognise a
competent request that merely falls outside scope (a $49.0\%$ false-approved
rate). Adding the blocklist \emph{on top of} the allowlist does not improve on
the allowlist alone: raw detection edges up, but the benign false-blocked rate
roughly doubles ($3.6\%\to8.2\%$) and overall accuracy slips
($95.9\%\to94.9\%$). The iterated allowlist thus remains the strongest
configuration: the allowlist is both necessary and sufficient to enforce the
functional boundary, and a blocklist neither substitutes for it nor usefully
complements it.

\begin{table*}[t]
\centering
\small
\caption{Overall performance of Persona Guardrail on \bench across guardrail
configurations (no mixed intents, percentages; i.e.\ the benign/adversarial/OOD
setting). Best value per row in \textbf{bold}.}
\label{tab:overall_rates}
\begin{tabular}{lccccc}
\toprule
\textbf{Metric (\%)} & \textbf{Baseline} & \textbf{+Allowlist} & \textbf{+Iterated} & \textbf{Blocklist} & \textbf{Allow+Block} \\
\midrule
Accuracy $\uparrow$                    & 85.7 & 93.2 & \textbf{95.9} & 71.2 & 94.9 \\
Precision $\uparrow$                   & 95.3 & 96.0 & \textbf{96.3} & 85.4 & 92.2 \\
Recall $\uparrow$                      & 75.0 & 90.1 & 95.3 & 51.0 & \textbf{98.0} \\
F1 $\uparrow$                          & 83.9 & 93.0 & \textbf{95.8} & 63.9 & 95.0 \\
False-approved rate (FAR) $\downarrow$ & 25.0 & 9.9  & 4.7  & 49.0 & \textbf{2.0} \\
False-blocked rate (FBR) $\downarrow$  & 3.7  & 3.7  & \textbf{3.6} & 8.7  & 8.2 \\
\bottomrule
\end{tabular}
\end{table*}

\subsection{Per-Label Analysis}
\label{sec:per_label}

Table~\ref{tab:per_label} decomposes performance by request label, and
Table in Appendix~\ref{sec:appendix} breaks the strongest configuration (iterated
allowlist) down by agent. Two observations stand out. First, adversarial
detection is high for every allowlist configuration ($\geq 96.8\%$), and even
the description-only baseline and the generic blocklist catch most overt attacks
($92.6\%$ and $87.2\%$): explicit attacks are the easy case. Second,
out-of-domain detection is where the functional boundary matters most---it
climbs from $57.3\%$ (baseline) to $83.3\%$ (allowlist) to $93.5\%$ (iterated),
whereas the blocklist manages only $14.5\%$, confirming that identifying
competent-but-out-of-scope requests---not attack coverage---is the hard part of
the problem, and that only an allowlist addresses it. Per agent, the lower Slack
and Workspace OOD rates reflect the difficulty of the fluent, task-like
cross-persona OOD cases described in Section~\ref{sec:data_diversity}.


\begin{table}[t]
\centering
\small
\setlength{\tabcolsep}{3pt}
\caption{Per-label performance across configurations on \bench (no mixed intents, percentages). For benign we report the false-blocked rate (lower is
better); for adversarial and out-of-domain we report the detection rate
(higher is better).}
\label{tab:per_label}

\begin{tabular}{lccccc}
\toprule
\textbf{Label} &
\textbf{Base} &
\textbf{Allow} &
\textbf{Iter.} &
\textbf{Block} &
\textbf{Allow+Block} \\
\midrule
Benign (FBR $\downarrow$)      & 3.7  & 3.7  & 3.6  & 8.7  & 8.2  \\
Adversarial ($\uparrow$)        & 92.6 & 96.8 & 97.0 & 87.2 & 99.2 \\
Out-of-domain ($\uparrow$)      & 57.3 & 83.3 & 93.5 & 14.5 & 96.8 \\
\bottomrule
\end{tabular}
\end{table}

    \subsection{Ablation: Mixed Intents}
\label{sec:mixed_ablation}

The experiments above use the three-label setting (benign, adversarial,
out-of-domain). We additionally study \emph{mixed} turns---turns that combine a
valid in-scope request with an out-of-scope or adversarial rider---which raise a
labelling question: should a turn that contains \emph{some} legitimate intent be
approved (allowlist-presence semantics) or blocked (because it also carries
off-policy content)? We evaluate both labelling policies against the
no-mixed setting on the iterated-allowlist configuration
(Table~\ref{tab:mixed_ablation}).

The guardrail blocks $92.2\%$ of mixed turns in both policies (its predictions
do not depend on the label). Consequently, scoring mixed turns as
\emph{should-approve} inflates the false-blocked rate to $17.1\%$ and drops
accuracy to $88.6\%$, whereas scoring them as \emph{should-block} keeps accuracy
at $95.6\%$ with a $3.6\%$ false-blocked rate. In other words, the guardrail
behaves as if a mixed turn is off-policy, which is consistent with a strict
functional boundary that rejects any turn carrying a disallowed component.

\begin{table}[t]
\centering
\small
\caption{Mixed-intent ablation on the iterated-allowlist configuration
(percentages). ``No mixed'' excludes mixed turns; the other two columns include
them, labelled as should-approve or should-block, respectively. FAR/FBR are the
false-approved / false-blocked rates.}
\label{tab:mixed_ablation}
\setlength{\tabcolsep}{5pt}
\begin{tabular}{lccc}
\toprule
\textbf{Metric (\%)} & \textbf{No mixed} &
\begin{tabular}[c]{@{}c@{}}\textbf{Mixed =}\\\textbf{allowed}\end{tabular} &
\begin{tabular}[c]{@{}c@{}}\textbf{Mixed =}\\\textbf{blocked}\end{tabular} \\
\midrule
Accuracy $\uparrow$        & 95.9 & 88.6 & 95.6 \\
Precision $\uparrow$       & 96.3 & 82.5 & 96.9 \\
Recall $\uparrow$          & 95.3 & 95.3 & 94.8 \\
F1 $\uparrow$              & 95.8 & 88.4 & 95.8 \\
FAR $\downarrow$           & 4.7  & 4.7  & 5.2  \\
FBR $\downarrow$           & 3.6  & 17.1 & 3.6  \\
Mixed turns blocked        & --   & 92.2 & 92.2 \\
\bottomrule
\end{tabular}
\end{table}

\section{Production Deployment}

  While the previous sections focused on the design and evaluation of
  Persona Guardrail, deploying guardrails in production introduces a
  distinct set of engineering challenges. Production agentic systems operate
  across multiple agents with heterogeneous capabilities, evolving product
  requirements, and dynamic user behavior. Consequently, guardrails must not
  only provide strong security but also remain observable, maintainable, and
  adaptable throughout their lifecycle, while operating within the strict
  latency budget imposed by their position on the critical path.

  \subsection{Model Serving Infrastructure}

  The guardrail model---a 30B-parameter Qwen3 mixture-of-experts model that
  activates approximately 3B parameters per token and is quantized to
  FP8---is served with vLLM on NVIDIA Dynamo. Inference is disaggregated into
  a CPU frontend and GPU workers: a prefix-cache-aware router dispatches
  each request to a worker, which performs the model inference on a
  dedicated accelerator. The frontend and worker pools scale independently
  (Figure~\ref{fig:serving-arch} in the Appendix).

  Because the guardrail lies on the critical path of every interaction, the
  serving infrastructure is optimized for predictable latency and high
  availability. The model context length is capped to bound the per-request
  KV-cache footprint and worst-case inference latency. Workers register with
  a per-region discovery mechanism, and requests are routed only to healthy
  workers in the same region. Individual worker failures therefore reduce
  local capacity without interrupting the remaining workers, while regional
  isolation contains region-wide failures and permits application-level
  failover.

  \subsection{Serving Within a Latency Budget}

  Because a guardrail runs on every request and response, its latency is
  judged not in isolation but against a \emph{transparency budget}: it may
  consume only a small fraction of the host application's end-to-end
  processing time before users perceive it. We express this budget as a
  fixed fraction of application latency, yielding a sub-100\,ms P90 target
  for each guardrail model call.

  Meeting this target is made tractable by three properties of the guardrail
  workload.

  \noindent 1) \textbf{Single-token, prefill-dominated inference.}
  The guardrail is a binary classifier that emits one decision token
  (allow or block). Token generation is negligible, so latency is dominated
  by prompt prefill rather than decoding.

  \noindent 2) \textbf{A large, shared prompt prefix.}
  All requests share a static preamble containing the instructions and
  policy context. This preamble accounts for approximately 93\% of input
  tokens in our workload. Reusing its cached state reduces per-request
  prefill to the shorter, variable user request or agent response.

  \noindent 3) \textbf{Bounded context with cheap rejection.}
  The small fraction of prompts that exceed the context cap (2.1\% of
  production traffic) is rejected before GPU work is scheduled. This
  prevents oversized inputs from consuming capacity or inflating latency for
  the remaining traffic (Appendix~\ref{sec:prompt-size}).

  Beyond exploiting workload structure, we reduce per-call cost through
  model selection rather than compression alone. A mixture-of-experts model
  with only a few billion active parameters, served in FP8, provides the
  capacity of a larger model at the inference cost of a substantially
  smaller one. Together, disaggregated serving, prefix-cache reuse, bounded
  contexts, and one-token output target the prefill-dominated structure of
  the workload rather than generic token-generation speed.

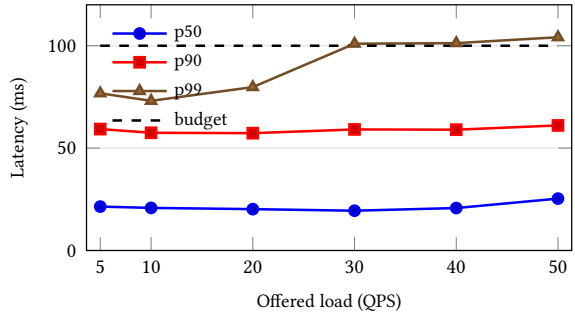
\begin{figure}[t]
  \centering
  \resizebox{0.9\columnwidth}{!}{%
  \begin{tikzpicture}
  \begin{axis}[
    width=0.82\linewidth,
    height=4.3cm,
    xlabel={Offered load (QPS)},
    ylabel={Latency (ms)},
    xmin=5, xmax=50,
    ymin=0, ymax=120,
    xtick={5,10,20,30,40,50},
    legend pos=north west,
    legend style={
      font=\scriptsize,
      draw=none,
      fill=none,
      cells={anchor=west}
    },
    tick label style={font=\scriptsize},
    label style={font=\scriptsize},
    every axis plot/.append style={line width=0.8pt},
    ymajorgrids,
    grid style={gray!20},
    enlarge x limits=0.03,
    enlarge y limits=false,
  ]
  \addplot+[mark=*, mark size=1.6pt] coordinates {
    (5,21.44) (10,20.80) (20,20.19)
    (30,19.43) (40,20.73) (50,25.34)
  };
  \addplot+[mark=square*, mark size=1.6pt] coordinates {
    (5,59.30) (10,57.49) (20,57.30)
    (30,59.12) (40,59.00) (50,61.08)
  };
  \addplot+[mark=triangle*, mark size=1.8pt] coordinates {
    (5,76.77) (10,73.09) (20,79.80)
    (30,100.99) (40,101.27) (50,104.15)
  };
  \addplot[dashed, thick, domain=5:50] {100};
  \legend{p50,p90,p99,budget}
  \end{axis}
  \end{tikzpicture}%
  }
  \caption{Guardrail-call latency versus offered load on a single
  accelerator with a 4{,}096-token context cap. P90 remains below the
  100\,ms objective throughout the tested range, while P99 crosses the
  objective at 30~QPS.}
  \label{fig:latency-qps}
  \end{figure}

  Under this design, the guardrail sustains its latency budget across the
  tested operating range (Figure~\ref{fig:latency-qps}). Between 5 and
  50~QPS, median latency ranges from 19.43 to 25.34\,ms and P90 latency from
  57.30 to 61.08\,ms. P99 increases from 76.77\,ms at 5~QPS to 104.15\,ms at
  50~QPS, crossing the 100\,ms objective at 30~QPS. These measurements are
  single-accelerator model-call latencies with a 4{,}096-token context cap;
  they do not include application or network overhead. Accelerator
  utilization remains below 65\% over this range. Relative to the prior
  general-purpose serving stack, the disaggregated design reduces
  guardrail-call latency by approximately 30\%.

  \subsection{Observability and Monitoring}

  Guardrail failures are frequently semantic rather than deterministic, so
  infrastructure health alone is insufficient. Persona Guardrail records
  telemetry for every processed interaction, including the decision, agent
  identifier, evaluated boundary, policy version, inference latency, and
  final outcome. Agent-level monitoring tracks allow and block rates and
  latency distributions, while operational monitoring measures throughput,
  resource utilization, availability, context-limit rejection, and
  end-to-end latency.

  Allow and block rates, together with disagreements between shadow and
  active policies, identify examples for investigation but are not
  themselves estimates of false-positive or false-negative rates. Such rates
  require examples with known labels or independently adjudicated samples.
  This distinction prevents agreement with the active policy from being
  mistaken for ground truth.

  Routine metrics contain identifiers and aggregate measurements rather
  than raw prompts, responses, or intent lists. Content retained for
  debugging or refinement is sampled, minimized or redacted,
  access-controlled, and deleted according to a bounded retention policy.
  These monitoring signals support the continuous guardrail evolution
  framework described in Section~\ref{sec:evolution}.

  \subsection{Guardrail Versioning and Safe Deployment}

  Unlike static security policies, guardrail specifications evolve as agent
  capabilities and attack techniques change. Persona Guardrail maintains an
  independent policy for each deployed agent, containing allowed and blocked
  intents for both user requests and agent responses. The same classifier
  can serve multiple agents, but each evaluation uses the intent lists
  associated with the current agent and interaction boundary. Candidate
  updates must satisfy predefined thresholds on security effectiveness,
  utility, reliability, and latency before deployment.

  Operationally, the editable intent taxonomy is separated from the policy
  enforced in production. Applying a persona to a policy copies its four
  intent lists into a deployment snapshot; later edits to the source persona
  do not affect that snapshot. A persona used by a shadow or active policy
  is not edited in place. Instead, it is cloned, the clone is refined, and
  the resulting candidate proceeds through offline evaluation, review, and
  rollout. Thus, each iteration creates an explicit candidate while the
  currently active allowlist remains unchanged.

  To reduce overfitting to a fixed benchmark, offline evaluation uses three
  complementary pools: a \emph{golden} set of curated, unambiguous behaviors
  that acts as a regression floor; an \emph{adversarial} set of prompt
  injections, jailbreaks, and difficult boundary cases; and a
  \emph{production} set sampled from live traffic. Examples used to motivate
  a refinement are kept separate from the held-out examples used to
  validate it. A candidate is accepted only if it satisfies the predefined
  thresholds across all three pools.

  Once approved, the candidate snapshot is distributed as declarative
  configuration rather than as a serving-code deployment. This allows
  per-agent policies to be promoted independently without redeploying the
  shared model-serving stack. Every production activation is gated by
  independent human review.

  New policies are initially deployed in shadow mode, where they evaluate
  sampled production traffic without affecting user-visible decisions.
  Their predictions are compared with those of the active policy to
  identify disagreements for analysis. A candidate that passes shadow
  evaluation is introduced through a limited enforcing rollout before
  becoming active. The same snapshot is used throughout these stages. If a
  regression is detected, the rollout is stopped and the previously active
  snapshot remains in effect or is restored.

  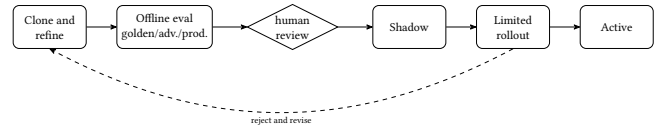
\begin{figure}[t]
  \centering
  \resizebox{\columnwidth}{!}{%
  \begin{tikzpicture}[
    font=\small,
    >={Stealth[round]},
    stage/.style={
      draw,
      rounded corners,
      minimum height=10mm,
      minimum width=17mm,
      align=center
    },
    gate/.style={
      draw,
      diamond,
      aspect=2,
      align=center,
      inner sep=1pt
    },
  ]
    \node[stage] (edit) {Clone and\\refine};
    \node[stage, right=7mm of edit] (eval)
      {Offline eval\\golden/adv./prod.};
    \node[gate, right=8mm of eval] (review)
      {human\\review};
    \node[stage, right=8mm of review] (shadow)
      {Shadow};
    \node[stage, right=7mm of shadow] (canary)
      {Limited\\rollout};
    \node[stage, right=7mm of canary] (active)
      {Active};

    \draw[->] (edit) -- (eval);
    \draw[->] (eval) -- (review);
    \draw[->] (review) -- (shadow);
    \draw[->] (shadow) -- (canary);
    \draw[->] (canary) -- (active);
    \draw[->, dashed] (canary.south) to[bend left=28]
      node[below, font=\scriptsize] {reject and revise}
      (edit.south);
  \end{tikzpicture}%
  }
  \caption{Lifecycle of a per-agent guardrail update. Each candidate is
  evaluated and reviewed before shadow and limited enforcement. A failed
  candidate leaves the previously active policy unchanged.}
  \label{fig:rollout}
  \end{figure}

  \subsection{Continuous Guardrail Evolution}
  \label{sec:evolution}

  Persona Guardrail evolves primarily by refining its human-readable intent
  taxonomy rather than retraining the classifier. Refinement may be
  triggered by newly introduced agent functionality, newly observed attacks,
  distribution shifts, or adjudicated classification errors. These signals
  initiate investigation; they never modify an active policy automatically.

  Examples are partitioned before refinement into a discovery pool and
  held-out validation pools. Correlated conversations, near-duplicate
  examples, and variants from the same attack campaign remain in the same
  partition. The discovery pool is used to propose changes, whereas
  acceptance is measured on held-out golden, adversarial, and
  production-derived examples. Deduplication, provenance tracking, and
  per-source sampling limits reduce evaluation leakage and the risk that one
  actor dominates a production-derived sample.
                                                                                                                                                 
  Golden and synthetic examples have labels assigned during construction.                                                                        
  For otherwise unlabeled production examples, a stronger judge model is                                                                         
  given the agent's functional boundary and the interaction, but not the                                                                         
  guardrail's decision, and predicts whether the interaction is permitted.                                                                       
  Judge-model disagreements with the guardrail, together with a sample of                                                                        
  agreements, are independently adjudicated before inclusion in reported                                                                         
  metrics. We treat an out-of-scope or adversarial interaction as the                                                                            
  positive class: a true positive is correctly blocked, a false negative is                                                                      
  incorrectly allowed, a false positive is a legitimate interaction that is                                                                      
  blocked, and a true negative is a legitimate interaction that is allowed.                                                                      
  Thus, disagreement with either the judge or the active policy is not by                                                                        
  itself treated as an error label.                                                                                                              
                                                                                                                                                 
  A separate offline proposer groups errors from the discovery pool and                                                                          
  suggests changes only to a new draft. The proposer cannot modify labels,                                                                       
  approve its proposal, or change production policy. The resulting candidate                                                                     
  is evaluated on the held-out pools and enters the review and rollout                                                                           
  lifecycle described above only if it satisfies the predefined security,                                                                        
  utility, reliability, and latency thresholds.                                                                                                  
                                                                                                                                                 
  
\subsection{Human-in-the-Loop Guardrail Management}

Although Persona Guardrail automates much of the guardrail lifecycle, human oversight remains essential for maintaining high-quality functional and security boundaries. Misclassified interactions identified through production monitoring are periodically reviewed by developers or security analysts to determine whether they correspond to evolving user intent, product changes, or novel attack strategies. Legitimate user interactions that fall outside the current allowlist may indicate that the agent's functional boundary should be expanded, whereas newly observed attack patterns may require updates to the blocklist.

Human review is particularly valuable for validating ambiguous interactions and resolving conflicts between usability and security. Rather than directly incorporating every production observation into the guardrail specification, Persona Guardrail treats human feedback as a high-confidence signal for refining both functional and security boundaries.

\subsection{Multi-Agent Deployment}

Large production systems often consist of numerous specialized agents, each serving a distinct purpose while sharing common infrastructure and security requirements. Deploying independent guardrails for every agent quickly becomes difficult to manage as the number of agents increases.

Persona Guardrail addresses this challenge by separating \textbf{agent-specific
functional boundaries} from \textbf{shared security knowledge}.
Per-agent functional boundaries are stored as versioned records in a relational
(MySQL) store and managed through the Registry, which exposes a management API
for creating, updating, and versioning each agent's allowlists and
out-of-domain definitions. The registry serves as the single source of truth
consumed by both the tuning pipeline and the policy engine, giving agent teams a
self-serve control plane while ensuring a consistent view of policy across the
fleet. The shared security blocklist---covering common attack patterns such as
prompt injection, jailbreaks, indirect prompt injection, and prompt
leakage---is maintained centrally and referenced by every agent, so a signature
added once is enforced everywhere.

This separation enables new attack signatures discovered in one agent to be propagated rapidly across the entire deployment while allowing each agent to maintain its own functional specification. As a result, Persona Guardrail scales efficiently to heterogeneous multi-agent deployments without requiring every guardrail to be developed and maintained independently.

\subsection{Results and Operational Lessons}


In production, Persona Guardrail meets its latency budget with headroom while sustaining a false-block rate below 1\% and a false-allow rate below 2\% across various agents.

Our deployment experience highlights that maintaining production guardrails is fundamentally an ongoing operational process rather than a one-time model development task. Effective guardrails require continuous observability, standardized evaluation, structured feedback collection, and disciplined deployment practices. By integrating these operational components into a unified lifecycle, Persona Guardrail enables organizations to maintain robust security while adapting to evolving products, user behavior, and adversarial threats.

Two further lessons stand out from operating Persona Guardrail at scale. First, a guardrail is a specialized inference workload---prefill-dominated and single-token, with a large shared prompt prefix---rather than a general-purpose chat model; meeting its transparency budget followed from exploiting that structure through prefix-cache reuse, frontend/decode disaggregation, and an efficient model, not from generic serving tuning. Second, decoupling policy iteration from infrastructure deployment---by expressing guardrail versions as peer-reviewed configuration changes---was essential to reconciling rapid iteration with production safety.

\paragraph{Scope and Limitations}

Persona Guardrail is a defense-in-depth control evaluated for direct malicious-user and out-of-domain interactions under the defined threat model. The reported results do not establish complete protection against indirect prompt injection, compromised components, unauthorized intermediate tool actions, multi-turn attacks, or other threats outside that model. Deployment decisions and acceptable error rates remain application-specific.

\section{Related Work}
Integrating LLMs with external tools, APIs etc and deploying it to serve specific a purpose introduces attack
surfaces beyond traditional LLM safety, including prompt injection, indirect
prompt injection, goal hijacking, retrieval poisoning, and tool
misuse~\cite{greshake2023prompt, liu2024promptinjection}. A recent
systematization of knowledge~\cite{kim2026agenticsok} consolidates these attacks
and defenses and highlights needs for scalable, adaptive, and deployable
runtime protection---the gap Persona Guardrail targets.

\paragraph{Security Defenses for \AI{} Agents} Runtime defenses target specific attacks during execution. Task
Shield~\cite{jia2025taskshield} enforces \emph{task alignment}, checking that
intermediate steps serve the user's task; MELON~\cite{zhu2025melon} re-executes
the trajectory with a masked, task-neutral user prompt and flags injection when
the masked and original tool calls coincide; and
LlamaFirewall~\cite{chennabasappa2025llamafirewall}, in production at Meta, chains
jailbreak, chain-of-thought alignment, and code scanners over agent inputs,
reasoning, and outputs. A complementary line enforces security by
construction: CaMeL~\cite{debenedetti2025camel} separates control and data flow
so untrusted content cannot alter the program; the Instruction
Hierarchy~\cite{wallace2024instruction}, StruQ~\cite{chen2025struq}, and
SecAlign~\cite{chen2025secalign} train models to prioritize trusted instructions;
and Progent~\cite{shi2025progent} and IsolateGPT~\cite{wu2025isolategpt} constrain
tool privileges and isolate components. These defenses chiefly address
\emph{indirect}, environment-mediated attacks on the agent's internal execution.
Persona Guardrail is complementary: under a \emph{direct malicious-user} threat
model it models an agent's \textbf{functional boundary} and enforces it with
input and output guardrails, capturing adversarial, out-of-domain, and
policy-violating turns alike.

\paragraph{Guardrails for LLM Applications} Application guardrails fall into rule-based toolkits
(NeMo~\cite{rebedea2023nemo}), model-based classifiers (Llama
Guard~\cite{inan2023llamaguard}, ShieldGemma~\cite{zeng2024shieldgemma},
WildGuard~\cite{han2024wildguard}), and hybrids. Closest to our work,
Constitutional Classifiers~\cite{sharma2025constitutional} train input and output
classifiers from a natural-language policy with production-grade overhead. These
are typically static, inference-time classifiers over \emph{global} harm
categories. Persona Guardrail instead enforces a \emph{per-agent functional
boundary}---an allowlist of intended intents that also rejects competent but
out-of-scope requests---and treats guardrails as \textbf{continuously evolving
systems} maintained via specification, standardized evaluation, and
feedback-driven refinement.

\paragraph{Benchmarks for Agentic Security} Agentic-security evaluation has standardized around
benchmarks for attacks and utility---AgentDojo~\cite{debenedetti2025agentdojo},
InjecAgent~\cite{zhan2024injecagent}, AgentHarm~\cite{agentharm2024}, and Agent
Security Bench~\cite{zhang2025asb}---none of which label whether a request lies
inside an agent's intended function. Recognizing out-of-scope requests has a long
lineage in task-oriented dialog, e.g., CLINC150~\cite{larson2019clinc}. PAGE
builds on AgentDojo's domains but adds agent-specific out-of-domain data and
labels for both user and agent turns, jointly measuring attack robustness and
rejection of out-of-scope requests.

In sum, prior work advances attacks
and class-specific runtime, architectural, and training defenses, plus
classifiers for global safety policies; Persona Guardrail contributes a
production-grade framework for specifying, enforcing, evaluating, and
continuously evolving an agent's functional boundary under a direct
malicious-user threat model.

\section{Conclusion}
We presented Persona Guardrail, a runtime defense framework that enforces agent-specific functional boundaries through synchronous validation of user inputs and final agent responses. We also introduced PAGE, a benchmark covering benign, adversarial, and out-of-domain interactions across user and agent turns. On the evaluated non-mixed subset of PAGE, an iteratively refined allowlist improved accuracy from 85.7\% for a generic LLM judge to 95.9\%, increased out-of-domain detection from 57.3\% to 93.5\%, and reduced the false-approved rate from 25.0\% to 4.7\%. Our deployment study further shows that functional-boundary classification can be served within a sub-100\,ms P90 per-call latency budget using a disaggregated, prefix-cache-aware inference architecture. These results indicate that explicit, agent-specific allowlists can complement attack-focused defenses by rejecting both malicious requests and requests outside an agent's intended scope.

Future work will extend Persona Guardrail in two directions. First, we plan to develop conversation-level persona enforcement that evaluates each interaction in the context of previous turns, enabling the guardrail to identify functional-boundary violations that emerge across a multi-turn conversation. Second, we plan to integrate persona enforcement at the model execution layer so that intermediate tool calls can be validated against the agent's functional boundary before execution. Together, these extensions will expand Persona Guardrail from turn-level input and output validation to conversation- and tool-level enforcement.

\bibliographystyle{ACM-Reference-Format}
\bibliography{ref}

@inproceedings{kim2026agenticsok,
  author    = {Juhee Kim and Wenbo Guo and Dawn Song},
  title     = {{SoK}: Attack and Defense Landscape of Agentic {AI} Systems},
  booktitle = {35th USENIX Security Symposium (USENIX Security 26)},
  year      = {2026},
  pages     = {6047--6066},
  publisher = {USENIX Association},
  url       = {https://www.usenix.org/conference/usenixsecurity26/presentation/kim-juhee-agentic}
}

@article{yao2023react,
  title={React: Synergizing reasoning and acting in language models},
  author={Yao, Shunyu and Zhao, Jeffrey and Yu, Dian and Du, Nan and Shafran, Izhak and Narasimhan, Karthik and Cao, Yuan},
  journal={arXiv preprint arXiv:2210.03629},
  year={2022}
}

@inproceedings{schick2023toolformer,
  author    = {Timo Schick and Jane Dwivedi-Yu and Roberto Dess{\`i} and Roberta Raileanu and Maria Lomeli and Eric Hambro and Luke Zettlemoyer and Nicola Cancedda and Thomas Scialom},
  title     = {Toolformer: Language Models Can Teach Themselves to Use Tools},
  booktitle = {Advances in Neural Information Processing Systems (NeurIPS)},
  volume    = {36},
  year      = {2023},
  url       = {https://proceedings.neurips.cc/paper_files/paper/2023/hash/d842425e4bf79ba039352da0f658a906-Abstract-Conference.html}
}

@article{wu2024autogen,
  title={Autogen: Enabling next-gen llm applications via multi-agent conversation},
  author={Wu, Qingyun and Bansal, Gagan and Zhang, Jieyu and Wu, Yiran and Li, Beibin and Zhu, Erkang and Jiang, Li and Zhang, Xiaoyun and Zhang, Shaokun and Liu, Jiale and others},
  journal={arXiv preprint arXiv:2308.08155},
  year={2023}
}

@misc{owasp2025llm,
  title        = {OWASP Top 10 for LLM Applications 2025},
  author       = {{OWASP Foundation}},
  year         = {2025},
  howpublished = {\url{https://genai.owasp.org/llm-top-10/}},
  note         = {Version 2.0}
}

@inproceedings{jia2025taskshield,
  title={The task shield: Enforcing task alignment to defend against indirect prompt injection in llm agents},
  author={Jia, Feiran and Wu, Tong and Qin, Xin and Squicciarini, Anna},
  booktitle={Proceedings of the 63rd Annual Meeting of the Association for Computational Linguistics (Volume 1: Long Papers)},
  pages={29680--29697},
  year={2025}
}

@inproceedings{greshake2023prompt,
  title={Not what you've signed up for: Compromising real-world llm-integrated applications with indirect prompt injection},
  author={Greshake, Kai and Abdelnabi, Sahar and Mishra, Shailesh and Endres, Christoph and Holz, Thorsten and Fritz, Mario},
  booktitle={Proceedings of the 16th ACM workshop on artificial intelligence and security},
  pages={79--90},
  year={2023}
}

@misc{promptshield,
  author       = {{Microsoft}},
  title        = {Prompt Shields for Azure AI Content Safety},
  year         = {2025},
  howpublished = {\url{https://learn.microsoft.com/azure/ai-services/content-safety/concepts/jailbreak-detection}}
}

@inproceedings{agentharm2024,
  title={Agentharm: A benchmark for measuring harmfulness of llm agents},
  author={Andriushchenko, Maksym and Souly, Alexandra and Dziemian, Mateusz and Duenas, Derek and Lin, Maxwell and Wang, Justin and Hendrycks, Dan and Zou, Andy and Kolter, Zico and Fredrikson, Matt and others},
  booktitle={International Conference on Learning Representations},
  volume={2025},
  pages={79185--79220},
  year={2025}
}

@article{debenedetti2025agentdojo,
  title={Agentdojo: A dynamic environment to evaluate prompt injection attacks and defenses for llm agents},
  author={Debenedetti, Edoardo and Zhang, Jie and Balunovic, Mislav and Beurer-Kellner, Luca and Fischer, Marc and Tram{\`e}r, Florian},
  journal={Advances in neural information processing systems},
  volume={37},
  pages={82895--82920},
  year={2024}
}

@article{qwen3vl2025,
  title        = {Qwen3-VL Technical Report},
  author       = {Bai et al.},
  journal      = {arXiv preprint arXiv:2511.21631},
  year         = {2025},
  eprint       = {2511.21631},
  archivePrefix= {arXiv},
  primaryClass = {cs.CL},
  url          = {https://arxiv.org/abs/2511.21631}
}

@inproceedings{liu2024promptinjection,
  title={Formalizing and benchmarking prompt injection attacks and defenses},
  author={Liu, Yupei and Jia, Yuqi and Geng, Runpeng and Jia, Jinyuan and Gong, Neil Zhenqiang},
  booktitle={33rd USENIX Security Symposium (USENIX Security 24)},
  pages={1831--1847},
  year={2024}
}

@inproceedings{zhu2025melon,
  author    = {Kaijie Zhu and
               Xianjun Yang and
               Jindong Wang and
               Wenbo Guo and
               William Yang Wang},
  title     = {{MELON}: Provable Defense Against Indirect Prompt Injection Attacks in {AI} Agents},
  booktitle = {Proceedings of the 42nd International Conference on Machine Learning (ICML)},
  series    = {Proceedings of Machine Learning Research},
  volume    = {267},
  year      = {2025},
  eprint    = {2502.05174},
  archivePrefix = {arXiv},
  primaryClass  = {cs.CR},
  url       = {https://arxiv.org/abs/2502.05174}
}

@misc{chennabasappa2025llamafirewall,
  author    = {Sahana Chennabasappa and
               Cyrus Nikolaidis and
               Daniel Song and
               David Molnar and
               Stephanie Ding and
               Shengye Wan and
               Spencer Whitman and
               Lauren Deason and
               Nicholas Doucette and
               Abraham Montilla and
               Alekhya Gampa and
               Beto de Paola and
               Dominik Gabi and
               James Crnkovich and
               Jean-Christophe Testud and
               Kat He and
               Rashnil Chaturvedi and
               Wu Zhou and
               Joshua Saxe},
  title     = {{LlamaFirewall}: An Open Source Guardrail System for Building Secure {AI} Agents},
  year      = {2025},
  eprint    = {2505.03574},
  archivePrefix = {arXiv},
  primaryClass  = {cs.CR},
  url       = {https://arxiv.org/abs/2505.03574}
}

@misc{debenedetti2025camel,
  author    = {Edoardo Debenedetti and
               Ilia Shumailov and
               Tianqi Fan and
               Jamie Hayes and
               Nicholas Carlini and
               Daniel Fabian and
               Christoph Kern and
               Chongyang Shi and
               Andreas Terzis and
               Florian Tram{\`e}r},
  title     = {Defeating Prompt Injections by Design},
  year      = {2025},
  eprint    = {2503.18813},
  archivePrefix = {arXiv},
  primaryClass  = {cs.CR},
  url       = {https://arxiv.org/abs/2503.18813}
}

@misc{wallace2024instruction,
  author    = {Eric Wallace and
               Kai Xiao and
               Reimar Leike and
               Lilian Weng and
               Johannes Heidecke and
               Alex Beutel},
  title     = {The Instruction Hierarchy: Training {LLMs} to Prioritize Privileged Instructions},
  year      = {2024},
  eprint    = {2404.13208},
  archivePrefix = {arXiv},
  primaryClass  = {cs.CR},
  url       = {https://arxiv.org/abs/2404.13208}
}

@inproceedings{chen2025struq,
  author    = {Sizhe Chen and
               Julien Piet and
               Chawin Sitawarin and
               David Wagner},
  title     = {{StruQ}: Defending Against Prompt Injection with Structured Queries},
  booktitle = {34th USENIX Security Symposium (USENIX Security '25)},
  year      = {2025},
  publisher = {USENIX Association},
  url       = {https://www.usenix.org/conference/usenixsecurity25/presentation/chen-sizhe}
}

@inproceedings{chen2025secalign,
  author    = {Sizhe Chen and
               Arman Zharmagambetov and
               Saeed Mahloujifar and
               Kamalika Chaudhuri and
               David Wagner and
               Chuan Guo},
  title     = {{SecAlign}: Defending Against Prompt Injection with Preference Optimization},
  booktitle = {Proceedings of the 2025 ACM SIGSAC Conference on Computer and Communications Security (CCS '25)},
  year      = {2025},
  doi       = {10.1145/3719027.3744836},
  url       = {https://arxiv.org/abs/2410.05451}
}

@misc{shi2025progent,
  author    = {Tianneng Shi and
               Jingxuan He and
               Zhun Wang and
               Hongwei Li and
               Linyu Wu and
               Wenbo Guo and
               Dawn Song},
  title     = {{Progent}: Progent: Securing AI Agents with Privilege Control},
  year      = {2025},
  eprint    = {2504.11703},
  archivePrefix = {arXiv},
  primaryClass  = {cs.CR},
  url       = {https://arxiv.org/abs/2504.11703}
}

@inproceedings{wu2025isolategpt,
  author    = {Yuhao Wu and
               Franziska Roesner and
               Tadayoshi Kohno and
               Ning Zhang and
               Umar Iqbal},
  title     = {{IsolateGPT}: An Execution Isolation Architecture for {LLM}-Based Agentic Systems},
  booktitle = {Proceedings of the Network and Distributed System Security Symposium (NDSS)},
  year      = {2025},
  doi       = {10.14722/ndss.2025.241131},
  url       = {https://www.ndss-symposium.org/ndss-paper/isolategpt-an-execution-isolation-architecture-for-llm-based-agentic-systems/}
}

@inproceedings{rebedea2023nemo,
  author    = {Traian Rebedea and
               Razvan Dinu and
               Makesh Narsimhan Sreedhar and
               Christopher Parisien and
               Jonathan Cohen},
  title     = {{NeMo Guardrails}: A Toolkit for Controllable and Safe {LLM} Applications with Programmable Rails},
  booktitle = {Proceedings of the 2023 Conference on Empirical Methods in Natural Language Processing: System Demonstrations (EMNLP)},
  year      = {2023},
  pages     = {431--445},
  publisher = {Association for Computational Linguistics},
  doi       = {10.18653/v1/2023.emnlp-demo.40},
  url       = {https://aclanthology.org/2023.emnlp-demo.40}
}

@misc{inan2023llamaguard,
  author    = {Hakan Inan and
               Kartikeya Upasani and
               Jianfeng Chi and
               Rashi Rungta and
               Krithika Iyer and
               Yuning Mao and
               Michael Tontchev and
               Qing Hu and
               Brian Fuller and
               Davide Testuggine and
               Madian Khabsa},
  title     = {{Llama Guard}: {LLM}-based Input-Output Safeguard for Human-{AI} Conversations},
  year      = {2023},
  eprint    = {2312.06674},
  archivePrefix = {arXiv},
  primaryClass  = {cs.CL},
  url       = {https://arxiv.org/abs/2312.06674}
}

@misc{zeng2024shieldgemma,
  author    = {Wenjun Zeng and
               Yuchi Liu and
               Ryan Mullins and
               Ludovic Peran and
               Joe Fernandez and
               Hamza Harkous and
               Karthik Narasimhan and
               Drew Proud and
               Piyush Kumar and
               Bhaktipriya Radharapu and
               Olivia Sturman and
               Oscar Wahltinez},
  title     = {{ShieldGemma}: Generative {AI} Content Moderation Based on Gemma},
  year      = {2024},
  eprint    = {2407.21772},
  archivePrefix = {arXiv},
  primaryClass  = {cs.CL},
  url       = {https://arxiv.org/abs/2407.21772}
}

@inproceedings{han2024wildguard,
  author    = {Seungju Han and
               Kavel Rao and
               Allyson Ettinger and
               Liwei Jiang and
               Bill Yuchen Lin and
               Nathan Lambert and
               Yejin Choi and
               Nouha Dziri},
  title     = {{WildGuard}: Open One-Stop Moderation Tools for Safety Risks, Jailbreaks, and Refusals of {LLMs}},
  booktitle = {Advances in Neural Information Processing Systems (NeurIPS) Datasets and Benchmarks Track},
  year      = {2024},
  eprint    = {2406.18495},
  archivePrefix = {arXiv},
  primaryClass  = {cs.CL},
  url       = {https://arxiv.org/abs/2406.18495}
}

@misc{sharma2025constitutional,
  author    = {Mrinank Sharma and
               Meg Tong and
               Jesse Mu and
               Jerry Wei and
               Jorrit Kruthoff and
               Scott Goodfriend and
               Euan Ong and
               others},
  title     = {Constitutional Classifiers: Defending against Universal Jailbreaks across Thousands of Hours of Red Teaming},
  year      = {2025},
  note      = {Safeguards Research Team, Anthropic},
  eprint    = {2501.18837},
  archivePrefix = {arXiv},
  primaryClass  = {cs.CL},
  url       = {https://arxiv.org/abs/2501.18837}
}

@inproceedings{zhan2024injecagent,
  author    = {Qiusi Zhan and
               Zhixiang Liang and
               Zifan Ying and
               Daniel Kang},
  title     = {{InjecAgent}: Benchmarking Indirect Prompt Injections in Tool-Integrated Large Language Model Agents},
  booktitle = {Findings of the Association for Computational Linguistics (ACL)},
  year      = {2024},
  doi       = {10.18653/v1/2024.findings-acl.624},
  url       = {https://aclanthology.org/2024.findings-acl.624/}
}

@inproceedings{zhang2025asb,
  author    = {Hanrong Zhang and
               Jingyuan Huang and
               Kai Mei and
               Yifei Yao and
               Zhenting Wang and
               Chenlu Zhan and
               Hongwei Wang and
               Yongfeng Zhang},
  title     = {Agent Security Bench ({ASB}): Formalizing and Benchmarking Attacks and Defenses in {LLM}-based Agents},
  booktitle = {International Conference on Learning Representations (ICLR)},
  year      = {2025},
  eprint    = {2410.02644},
  archivePrefix = {arXiv},
  primaryClass  = {cs.CR},
  url       = {https://arxiv.org/abs/2410.02644}
}

@inproceedings{larson2019clinc,
  author    = {Stefan Larson and
               Anish Mahendran and
               Joseph J. Peper and
               Christopher Clarke and
               Andrew Lee and
               Parker Hill and
               Jonathan K. Kummerfeld and
               Kevin Leach and
               Michael A. Laurenzano and
               Lingjia Tang and
               Jason Mars},
  title     = {An Evaluation Dataset for Intent Classification and Out-of-Scope Prediction},
  booktitle = {Proceedings of the 2019 Conference on Empirical Methods in Natural Language Processing and the 9th International Joint Conference on Natural Language Processing (EMNLP-IJCNLP)},
  year      = {2019},
  doi       = {10.18653/v1/D19-1131},
  url       = {https://aclanthology.org/D19-1131}
}

\begin{table*}[t]
\centering
\small
\caption{Decision counts underlying Table~\ref{tab:overall_rates} ($2{,}000$ turns per configuration).}
\label{tab:overall_counts}
\begin{tabular}{lccccc}
\toprule
\textbf{Decision} & \textbf{Baseline} & \textbf{+Allowlist} & \textbf{+Iterated} & \textbf{Blocklist} & \textbf{Allow+Block} \\
\midrule
Correct blocked (TP)  & 747 & 897 & 949 & 508 & 976 \\
False blocked (FP)    & 37  & 37  & 36  & 87  & 82  \\
Correct approved (TN) & 967 & 967 & 968 & 917 & 922 \\
False approved (FN)   & 249 & 99  & 47  & 488 & 20  \\
\bottomrule
\end{tabular}
\end{table*}

\appendix 
\section{Additional Experimental Results}
\label{sec:appendix}

This appendix presents additional experimental results that complement the evaluation in Section~4. Table~\ref{tab:overall_counts} reports the underlying decision counts for each guardrail configuration. Table~\ref{tab:data_length} summarizes the length distribution of messages in the PAGE benchmark, illustrating the diversity of the evaluation corpus. 
Table~\ref{tab:per_agent_best} reports the performance of the iterated configuration for each of the four evaluated agent domains, demonstrating that Persona Guardrail generalizes consistently across agents. Table~\ref{tab:iteration_heldout} validates the iterative allowlist refinement on disjoint held-out splits, confirming the gains are not an artifact of tuning on the evaluation set.

\subsection{Iterating the allowlist on held-out data.}
To ensure refinement is not overfitting, we partitioned \bench into disjoint
two-way ($50/50$) and three-way ($33/33/33$) splits and scored iterations on
unseen data (Table~\ref{tab:iteration_heldout}). Testing \emph{v1} (initial),
\emph{v2} (iterated, Table~\ref{tab:overall_rates}), and \emph{v3} (refined from
v2's part~2 errors, tested on part~3) showed monotonic accuracy gains on unseen
data ($92.4\%\!\to\!96.0\%$ for $50/50$;
$91.9\%\!\to\!96.1\%\!\to\!96.7\%$ for $33/33/33$). Furthermore, \emph{v3} cut
false approvals ($5.1\%\!\to\!2.7\%$) and boosted out-of-domain detection
($93.9\%\!\to\!95.8\%$), confirming true functional-boundary gains over
overfitting.

\subsection{Defense against informed attackers}

In additional experiments, we tested how Persona Guard withstands automated attackers possessing varying levels of system knowledge. We evaluated several models as attackers (Opus 4.8, Opus 5, Sonnet 5, Luna, Fable 5, and GPT-5.5) across three knowledge tiers: general architecture, the exact model and prompt structure, and the verbatim allowed intent lists. Attackers were granted multiple turns to bypass the guardrail if their initial attempts failed.
We observed that access to the verbatim intent list provided the greatest advantage to the attacker, a scenario that carries relatively low real-world risk. Conversely, knowing the exact underlying model offered no additional benefit once the attacker was already aware of the guardrail's presence. Even in the worst-case scenario, where the attacker had full access to the intent list, the system blocked over 85\% of attacks generated by the most capable model, while maintaining a 90\%+ detection rate across all other scenarios.
A notable limitation during these experiments was the safety tuning of the closed-source models used as attackers. Because these models frequently refused to generate adversarial prompts, we observed a refusal rate between 17\% and 71\% on first-turn attacks.
\section{Distribution of Production Traffic}
\label{sec:prompt-size}

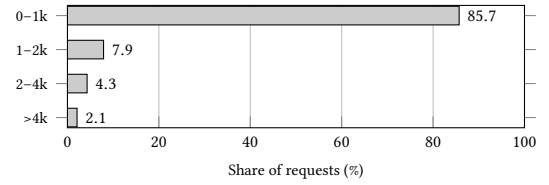
\begin{figure}[t]
\centering
\begin{tikzpicture}
\begin{axis}[
  xbar,
  width=0.9\columnwidth,
  height=3.2cm,
  xmin=0, xmax=100,
  xlabel={Share of requests (\%)},
  symbolic y coords={0--1k,1--2k,2--4k,>4k},,
  ytick=data,
  y dir=reverse,
  nodes near coords,
  nodes near coords style={font=\scriptsize},
  xmajorgrids,
  tick label style={font=\scriptsize},
  label style={font=\scriptsize},
  bar width=7pt,
]
\addplot [
    fill=black!20,
    draw=black
] coordinates {
  (85.7,0--1k)
  (7.9,1--2k)
  (4.3,2--4k)
  (2.1,>4k)
};
\end{axis}
\end{tikzpicture}
\caption{Prompt-size distribution of production guardrail traffic ($n{=}23{,}738$). Most requests are short (85.7\% under 1k tokens); only 2.1\% exceed the 4,096-token context cap and are rejected upstream.}
\label{fig:prompt-dist}
\end{figure}

As Figure~\ref{fig:prompt-dist} shows, most production guardrail prompts are short (under 1,000 tokens). The small fraction of requests that exceed the context limit are rejected before inference to protect latency and GPU utilization.

\begin{table}[t]
\centering
\small
\caption{Message length (words) by request kind, showing the wide and
kind-dependent spread that stresses guardrail robustness.}
\label{tab:data_length}
\begin{tabular}{lcccc}
\toprule
\textbf{Request kind} & \textbf{Median} & \textbf{Mean} & \textbf{P90} & \textbf{Max} \\
\midrule
Benign          & 17 & 20.7 & 33 & 121 \\
Adversarial     & 14 & 15.4 & 22 & 84 \\
Out-of-domain   & 23 & 30.9 & 63 & 121 \\
Mixed           & 37 & 44.1 & 68 & 152 \\
\midrule
\textbf{Overall}& 17 & 23.8 & 47 & 152 \\
\bottomrule
\end{tabular}
\end{table}

\begin{table}[t]
\centering
\small
\caption{Per-agent performance of the iterated-allowlist configuration on
\bench (no mixed intents, percentages). FBR is the benign false-blocked rate;
detection rates are the fraction of adversarial / out-of-domain turns correctly blocked.}
\label{tab:per_agent_best}
\begin{tabular}{lccc}
\toprule
\textbf{Agent} & \textbf{FBR $\downarrow$} & \textbf{Adv.\ det.\ $\uparrow$} & \textbf{OOD det.\ $\uparrow$} \\
\midrule
Workspace & 3.1 & 97.8 & 92.6 \\
Slack     & 4.0 & 98.4 & 85.6 \\
Banking   & 6.4 & 93.4 & 98.4 \\
Travel    & 0.8 & 98.3 & 97.6 \\
\midrule
\textbf{Average} & 3.6 & 97.0 & 93.5 \\
\bottomrule
\end{tabular}
\end{table}

\begin{table}[t]
\centering
\small
\setlength{\tabcolsep}{4pt}
\caption{Iterative allowlist refinement on disjoint held-out partitions of
\bench (percentages). Each iteration is scored on a partition
it did not inform. FAR/FBR are the false-approved / false-blocked rates; Adv/OOD
are adversarial / out-of-domain detection rates.}
\label{tab:iteration_heldout}
\begin{tabular}{lccccc}
\toprule
\textbf{Iteration (split, part)} & \textbf{Acc} & \textbf{FAR $\downarrow$} & \textbf{FBR $\downarrow$} & \textbf{Adv $\uparrow$} & \textbf{OOD $\uparrow$} \\
\midrule
v1 ($50/50$, p1)      & 92.4 & 11.0 & 4.2 & 96.5 & 81.5 \\
v2 ($50/50$, p2)      & 96.0 & 4.8  & 3.2 & 96.7 & 93.6 \\
\midrule
v1 ($33/33/33$, p1)   & 91.9 & 11.5 & 4.8 & 95.8 & 81.2 \\
v2 ($33/33/33$, p2)   & 96.1 & 5.1  & 2.7 & 95.8 & 93.9 \\
v3 ($33/33/33$, p3)   & 96.7 & 2.7  & 3.9 & 98.8 & 95.8 \\
\bottomrule
\end{tabular}
\end{table}

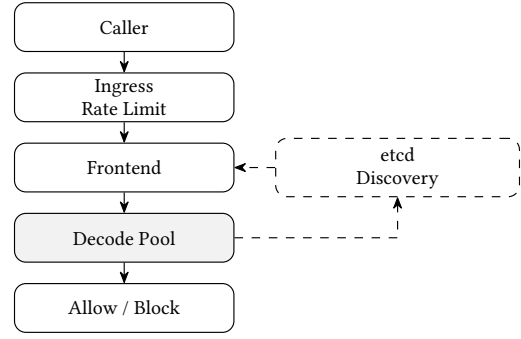
\begin{figure}[t]
\centering
\resizebox{0.8\columnwidth}{!}{%
\begin{tikzpicture}[
    node distance=2.5mm,
    font=\footnotesize,
    >={Stealth[round]},
    box/.style={
        draw,
        rounded corners,
        minimum width=27mm,
        minimum height=6mm,
        align=center,
        inner sep=2pt
    },
    gpu/.style={
        box,
        fill=black!5
    }
]

\node[box] (caller) {Caller};

\node[box, below=of caller] (mut)
{Ingress\\Rate Limit};

\node[box, below=of mut] (fe)
{Frontend};

\node[gpu, below=of fe] (dec)
{Decode Pool};

\node[box, below=of dec] (out)
{Allow / Block};

\node[
    draw,
    dashed,
    rounded corners,
    align=center,
    minimum width=30mm,
    minimum height=6mm,
    font=\footnotesize,
    right=5mm of fe
] (etcd)
{etcd\\Discovery};

\draw[->] (caller) -- (mut);
\draw[->] (mut) -- (fe);
\draw[->] (fe) -- (dec);
\draw[->] (dec) -- (out);

\draw[->, dashed] (etcd) -- (fe);
\draw[->, dashed] (dec) -| (etcd);

\end{tikzpicture}
}
\caption{Disaggregated serving architecture. A frontend performs prefix-cache-aware routing to a decode worker, which returns a single allow/block decision.}
\label{fig:serving-arch}
\end{figure}

\section{Open Science} 

We provide the following artifacts to support evaluation and reproduction of the paper's core results:
\begin{itemize}
    \item The PAGE benchmark, including the user- and agent-turn examples, persona and request-kind labels, and evaluation splits; and
    \item The agent-specific intent specifications used in the experiments, including both allowed and blocked intents for the four evaluated domains.
\end{itemize}
The artifacts are available through the following URL:
\url{https://github.com/uber-research/page-benchmark}.

We cannot release the exact production guardrail prompt due to security risks and proprietary details. Instead, the paper provides the methodology, model details, and intent specifications so researchers can build a comparable classifier. While exact reproduction isn't possible without our internal infrastructure, these tools allow researchers to achieve similar results.

\section{Use of Generative \AI{}}
We used generative \AI{} tools during this work for software development assistance
(including code generation and refactoring), generation of synthetic benchmark
examples and prompt templates, and editorial assistance for improving clarity,
grammar, and organization of the manuscript. All technical designs,
implementation decisions, experimental methodology, benchmark construction,
evaluation, and scientific conclusions were developed, verified, and approved by
the authors. All \AI{}-generated code, text, and synthetic data were reviewed and,
where appropriate, modified prior to inclusion in the final artifact.

\end{document}